\documentclass[a4paper]{article}
\usepackage{amsmath}
\usepackage{amssymb} %% symbols
\usepackage{amsthm} %% theorems

\usepackage[sort, round]{natbib} %% natbib
\usepackage{url} %% fixes potential escape characters in URLs

\usepackage{graphicx}
\usepackage{hyperref}
\usepackage[nolists]{endfloat}
\usepackage{dsfont} %% for expected value E

\graphicspath{{./fig/}}

\usepackage{tikz}
\usepackage[margin=1in]{geometry}

\usepackage[parfill]{parskip} %% no indents

\begin{document}
%\begin{frontmatter}

\title{Super-exponential bubbles in lab experiments: evidence for anchoring over-optimistic expectations on price}

\author{%
A.~H\"usler\\{\it{ahuesler@ethz.ch}} \\ETH Zurich \\Switzerland \and
D.~Sornette\\{\it{dsornette@ethz.ch}} \\ETH Zurich \\Switzerland \and
C.~H.~Hommes\\{\it{c.h.hommes@uva.nl}} \\University of Amsterdam \\ The Netherlands
}
\maketitle

%\author{A.~H\"usler}
%\ead{ahuesler@ethz.ch}
%\address{ETH Zurich }
%\author{D.~Sornette}
%\ead{dsornette@ethz.ch}
%\address{ETH Zurich }
%\author{C.~H.~Hommes}
%\ead{c.h.hommes@uva.nl}
%\address{University of Amsterdam}


\begin{abstract}
We analyze a controlled price formation experiment in the laboratory that 
shows evidence for bubbles. We calibrate two models that demonstrate
with high statistical significance that 
these laboratory bubbles have a tendency to grow faster than exponential
due to positive feedback. We show that the positive feedback operates
by traders continuously upgrading their over-optimistic expectations of future returns  
based on past prices rather than on realized returns. 
\end{abstract}

%\begin{keyword}
%rational expectations \sep financial bubbles \sep speculation \sep anchoring \sep
%laboratory experiments \sep behavioral model \sep super-exponential growth
%\sep positive feedback \sep behavioral expectations
%\end{keyword}


%\end{frontmatter}

% http://www.aeaweb.org/jel/jel_class_system.php

\textbf{Highlights:}
\begin{itemize}
%  \item We offer a behavioral model and interpretation of laboratory experiments that exhibit
%  financial bubbles.
%  \item We show that bubbles in controlled experiments can grow significantly faster than exponential.
%  \item We find that traders anchor their expectations more on price than on returns in these bubbles.
  \item We offer an interpretation of lab experiments that exhibit financial bubbles.
  \item We show that bubbles in controlled experiments can grow faster than exponential.
  \item We find traders anchor expectations more on price than on returns in these  bubbles.
\end{itemize}

%%%%%%%%%%%%%%%%%%%%%%%%%%%%%%%%%%%%%%%%%%%%%%%%%%%%%%%%%%%%%%%%%%%%%%%%%%%%%%%
%% header

%%%%%%%%%%%%%%%%%%%%%%%%%%%%%%%%%%%%%%%%%%%%%%%%%%%%%%%%%%%%%%%%%%%%%%%%%%%%%%%
%%%%%%%%%%%%%%%%%%%%%%%%%%%%%%%%%%%%%%%%%%%%%%%%%%%%%%%%%%%%%%%%%%%%%%%%%%%%%%%

%%%%%%%%%%%%%%%%%%%%%%%%%%%%%%%%%%%%%%%%%%%%%%%%%%%%%%%%%%%%%%%%%%%%%%%%%%%%%%%
%State the objectives of the work and provide an adequate background, avoiding a detailed literature survey or a summary of the results.
\section{Introduction}

Bubbles, defined as  significant persistent deviations from fundamental value,
express one of the most paradoxical behaviors of real financial markets.
Here, we analyze the dynamics of bubbles in a laboratory market (\cite{Hommes2008})
and focus on the regimes of strong deviations from the known fundamental values,
which we refer to as the bubble regimes.
Because this data is from a controlled environment, we can exclude exogenous influences such as news or private information.
We show that a model with exponential growth, corresponding to a constant rate of returns, cannot account for the 
observed transient explosive price increases. 
Models that incorporate positive feedback leading to faster-than-exponential growth are found to better describe the data. 

Research on financial bubbles has a rich literature (see e.g. \cite{Kaizoji2010}
for a recent review) aiming at explaining the origin of bubbles, their persistence and other properties. 
The theoretical literature has classified different type of bubbles. For instance, 
\cite{Blanchard1979} and \cite{Blanchard1982}
introduced rational expectation (RE) bubbles, i.e., bubbles that appear in the presence
of rational investors who are willing to earn the large returns offered during the duration
of the bubble as a remuneration for the risk that the bubble ends in a crash.
\cite{Tirole1982} argued that heterogeneous beliefs among traders is necessary for bubbles to develop.
\cite{DeLong1990} demonstrated that introducing noise traders in a universe of
rational speculators can amplify the size and duration of bubbles.
\cite{Brock1998} showed that endogenous switching between heterogeneous expectations rules, 
driven by their recent relative performance, generates bubble and crash dynamics of asset prices. 
\cite{Abreu2003} explained the persistence
of bubbles by the heterogeneous diagnostics of rational agents concerning the start time of the bubble,
which leads to a lack of  synchronization of their shorting of the underlying asset, and therefore prevents
them from stopping the bubble to blossom.
\cite{Lux2002} showed that the multiplicative stochastic process
proposed by Blanchard and Watson (1982), together with the no-arbitrage condition,
predicts a tail exponent of the distribution of returns smaller than $1$, which is incompatible
with empirical observations. 
\cite{Johansen1999} and \cite{Johansen2000} thus extended the Blanchard-Watson (1982) model of
RE bubbles by proposing models for the crash hazard rate that exhibit critical bifurcation points 
reflecting the imitation and herding behavior of the noise traders.
%\cite{Li2009} developed a model jointly specifying the price dynamics of  explosive RE bubbles together with mean-reverting residuals.
\cite{Gallegati2011} presented a model of bubbles and crashes, where crashes occur after a period of financial distress. 
\cite{Hommes2006} reviewed behavioral models of bubbles with fundamentalists trading against chartists.

\cite{Jarrow2007}, \cite{Jarrow2010} and \cite{Jarrow2011} developed local
martingale models of bubbles within the arbitrage-free martingale pricing technology that underlies
option pricing theory, based on the assumption that bubbles come together with
(or are defined by) a volatility growing faster than linearly with the
underlying price.
But \cite{Andersen2004}, among others, have shown that some 
(and perhaps most) bubbles are not associated
with an increase in volatility.
In particular,\cite{Bates1991} documented that the famous worldwide October 1987 crash
occurred at a minimum of the implied volatility, at least in the US\@.
% In contrast, a sharp rise of the historical as well as implied volatility is always observed {\it after} the crash ending the bubble \cite{Sornette1996}.
\cite{Guerkaynak2008} surveyed econometric tests of asset price bubbles and
showed that the econometric detection of asset price bubbles cannot be
achieved with a satisfactory degree of certainty: for each paper that finds evidence of
bubbles, there is another one that fits the data equally well without allowing for a bubble. 
%In this context, the originality of the present work is to introduce two models that, when calibrated
%to controlled laboratory market experiments, prove with high statistical significance 
%the existence of bubbles and that these laboratory bubbles have a tendency to grow faster than exponential
%due to positive feedback.

The present paper represents the first detailed quantitative calibration of simple models with positive feedback 
that unambiguously demonstrates the existence of positive feedback 
mechanisms and super-exponential bubbles in the price formation process.
It thus provides support within controlled laboratory set-ups
for the empirical evidence presented by Sornette et al.\ on historical
financial bubbles (see \cite{Jiang2010} and \cite{Kaizoji2010}\footnote{An extended version is
available at http://arxiv.org/abs/0812.2449} and references therein for an
overview).
%that include the worldwide bubble that ended with the Oct.
%1987 crash (\cite{Sornette1996}), the dot.com bubble that crashed in 2000 (\cite{Johansen2000b}), many bubbles in Latin-America and Asia (\cite{Johansen2001}), the real-estate bubbles in the UK (\cite{Zhou2003}) and the USA (\cite{Zhou2006}), the Oil bubble of 2008 (\cite{Sornette2009}),
%the Shanghai bubbles of 2007 and 2009 (\cite{Jiang2010}) and many others (\cite{Sornette2004} and \cite{Johansen2010}).

%%%%%%%%%%%%%%%%%%%%%%%%%%%%%%%%%%%%%%%%%%%%%%%%%%%%%%%%%%%%%%%%%%%%%%%%%%%%%%%
%Provide sufficient detail to allow the work to be reproduced. Methods already published should be indicated by a reference: only relevant modifications should be described.
\section{Material and methods}
	
In the experiment of \cite{Hommes2008}, participants (``traders'') were asked to forecast the price of a single asset in every turn.
The price of the asset evolves with the equation,
\begin{equation}
p_t = \frac{1}{1+r} \left[ \frac{1}{H}  \sum_{h=1}^H p_{t+1}^h+D \right],
\label{eq:marketPrice}
\end{equation}
where the market price $p_t$ at time $t$ is given as an average of the $H=6$ traders discounted price expectations;
$r=5\%$ is the interest rate, $p_{t+1}^h$ is the estimate of trader $h$ for the price for period $t+1$ based on information up to time $t-1$ and $D=3.00$ is the dividend.
Hence, today's price $p_t$ is simply the average of the current value of the traders' expectations for tomorrow $p_{t+1}^h$.
Note that the traders have to make a two period forecast; for their forecast $p^h_{t+1}$, only the prices up to time $t-1$ are available.

Traders are given the parameters above (but not the price forming \autoref{eq:marketPrice} itself) and are rewarded according to their prediction accuracy%
\footnote{The reward is proportional to the quadratic scoring rule $\max  \left\lbrace (1300 - 1300/49 (p_t - p_t^h))^2, 0 \right\rbrace $}.
The fundamental/equilibrium price $p^f$ (which traders could calculate) is $60$%
\footnote{$p^f = D/r = 3.00/5 \% = 60$}.
In our analysis, we focus on the realized price $p_t$ and not on the traders' individual estimates $p_t^h$.

Notwithstanding the existence of a clearly defined market price formula, this experiment is
remarkable in reporting realized prices that are quite loosely tied to the fundamental value,
because traders are rewarded more by correctly foreseeing the other traders' forecasts 
than by correctly calculating the fundamental price $p_f$.
Moreover, traders are allowed to estimate the asset value in a large price range between $0$ and $1000$
(where the upper bound is more than $16\times$ the fundamental value $p_f$).

%%%%%%%%%%%%%%%%%%%%%%%%%%%%%%%%%%%%%%%%%%%%%%%%%%%%%%%%%%%%%%%%%%%%%%%%%%%%%%%
%A Theory section should extend, not repeat, the background to the article already dealt with in the Introduction and lay the foundation for further work. In contrast, a Calculation section represents a practical development from a theoretical basis.
\section{Theory/calculation}
\label{sec:calc}

%%%%%%%%%%%%%%%%%%%%%%%%%%%%%%%%%%%%%%%
\subsection{Rational Bubble}

\cite{Hommes2008} discussed the rational bubble
\begin{equation}
p_t = (1+\hat{r})^t a_1 + b_1~,
\label{eq:rationalBubble}
\end{equation}
where $a_1$ is a positive constant.  This process fulfills the ``self-confirming'' nature of rational expectations
if the assumed interest rate $\hat{r}$ equals the interest rate $r$ from \autoref{eq:marketPrice} and $b_1$ equals the fundamental value of $p^f=60$%
\footnote{%
For a rational bubble, we have $\mathds{E}_t [ p_{t+1} + D ] =  c(1+r)^{t+1} + p^f (1+r) = (1+r)  p_t$.
%Further note that the growth rate is slightly increasing (i.e.\ super-exponential) for a very short time,
%but converges very fast and is bounded by $r$: $p_{t+1}^b / p_{t}^b - 1 \nearrow r $.
}.
In fact, \cite{Hommes2008} found that traders do use an interest rate $\hat{r}$ significantly larger than $r$ in four of the six groups and
hence their expectations are no longer rational (see \autoref{sec:results}).
Furthermore, the growth rate $\hat r$ is not constant, but is increasing as we will see later.

\cite{Todd2000} argued that ``decision-making agents in the real world must
arrive at their inferences using realistic amounts of time, information, and computational resources. [..] 
The most important aspects of an agent's environment are often created by the other agents it interacts with.''
Moreover, \cite{Tversky1974} presented
three heuristics that are employed in making judgments under uncertainty.
For our purposes, the heuristic that is relevant to interpret the groups' behaviors
is the ``adjustment from an \emph{anchor}, which is usually employed in numerical prediction
when a relevant value is available. These heuristics are highly economical and usually effective,
but they lead to systematic and predictable errors.'' (Emphasis is ours).

In the rest of this section, we are presenting two models  in which traders anchor their forecasts on
(1) price or (2) return.
Both models have in common that they can generate 
price growth that is significantly faster than exponential (as observed in the data)
and generalize the rational bubble of \autoref{eq:rationalBubble}.

%%%%%%%%%%%%%%%%%%%%%%%%%%%%%%%%%%%%%%%
\subsection{Anchoring on Price}

Generalizing the constant growth generated by \autoref{eq:rationalBubble}, 
we specify a model which allows faster or slower than exponential growth.
The growth rate $\log(\bar{p}_{t} / \bar{p}_{t-1})$ can be explained by the excess price $\bar{p}_{t-1}$
(which is the difference between the observed price $p_t$ and the fundamental price $p^f$) plus a constant:
\begin{equation}
	\log \left(\frac{\bar{p}_{t}}{\bar{p}_{t-1}} \right) = a_2 + b_2 \bar{p}_{t-1}. 
	\label{eq:regPrice}
\end{equation}
$a_2>0$ and $b_2>0$ would imply faster than exponential growth i.e.\ the growth rate grows itself.
For $b_2=0$, we recover the exponential growth (equivalent to the rational bubble \autoref{eq:rationalBubble} with $r=\hat{r}$).
We will see below that $b_2$ is typically significantly larger than zero,
indicating faster than exponential growth and positive feedback on the price.

One justification for the functional form (\autoref{eq:regPrice}) is that 
anchoring on price is commonly used in technical trading.
One of many patterns used are support and resistance levels which is nothing else but anchoring on price.
Although in violation with the efficient markets hypothesis, \cite{Lo2004} studied technical trading rules and found ``practical value'' for such technical rules.

%%%%%%%%%%%%%%%%%%%%%%%%%%%%%%%%%%%%%%%
\subsection{Anchoring on Return}
Alternatively, we check if the growth rate can be explained by the excess log-return $\log(\bar{p}_{t}/ \bar{p}_{t-1})$
following the following process
\begin{equation}
	\log \left(\frac{\bar{p}_{t+1}}{\bar{p}_{t}}\right) = a_3 + b_3 \log \left( \frac{\bar{p}_{t}}{\bar{p}_{t-1}} \right). 
	\label{eq:regReturn}
\end{equation}
The conditions that $a_3>0$ and $b_3>0$ implies again faster than exponential growth of the excess price $\bar{p}_t$ and positive feedback from past returns.
This model can be interpreted as a second order iteration or adaptive form of the exponential growth.

%%%%%%%%%%%%%%%%%%%%%%%%%%%%%%%%%%%%%%%%%%%%%%%%%%%%%%%%%%%%%%%%%%%%%%%%%%%%%%%
%Results should be clear and concise.
\section{Results}
\label{sec:results}

In this section, we estimate the parameters 
of the two processes and check for the statistical significance of $b_2$ and $b_3$
that express a positive feedback of price
(\autoref{eq:regPrice}) or of return (\autoref{eq:regReturn}) onto future returns.
In particular, we are interested in the lower 95\% confidence interval for the null
hypothesis that $b_2$ and $b_3$ are zero, to check for significant deviations that 
can confirm or not that price growth is indeed significantly faster than exponential
(which is the situation
corresponding to $b_2$ and $b_3$ greater than zero).
As the two models can be run over a multitude of different start and end points,
we present the results in graphical form instead of tables to provide better insight.

\cite{Hommes2008} identified bubbles in five out of the six groups.
Group one shows a somehow erratic price trajectory and no bubbles.
Groups five and six show some tendency towards bubbles, but the time horizon is too short for our analysis to get significant results.
Moreover, \cite{Hommes2008} found that the bubble in group five is compatible with the hypothesis of a rational bubble (\autoref{eq:rationalBubble}).
Hence, we focus on group number two, three and four.

\begin{center}
\begin{table}
\begin{tabular}{ccll}
Group & Time window & Description & Classification     \\
\hline
1 & NA & erratic price trajectory & ---                     \\
2 &  7 -- 26 & speculative bubble & anchoring on price      \\
3 &  7 -- 29 & speculative bubble & anchoring on price      \\
4 &  7 -- 21 & speculative bubble & anchoring on price      \\
5 & 29 -- 37 & rational bubble    & ---                     \\
6 & 23 -- 29 & speculative bubble & (too short for analysis)\\
\hline
\end{tabular}
\caption{Overview of bubbles reproduced from \cite{Hommes2008} with our own classification.}
\label{tab:overview}
\end{table}
\end{center}

%%%%%%%%%%%%%%%%%%%%%%%%%%%%%%%%%%%%%%%
\subsection{Group 2}

\begin{figure}[ht]
\centering
\includegraphics{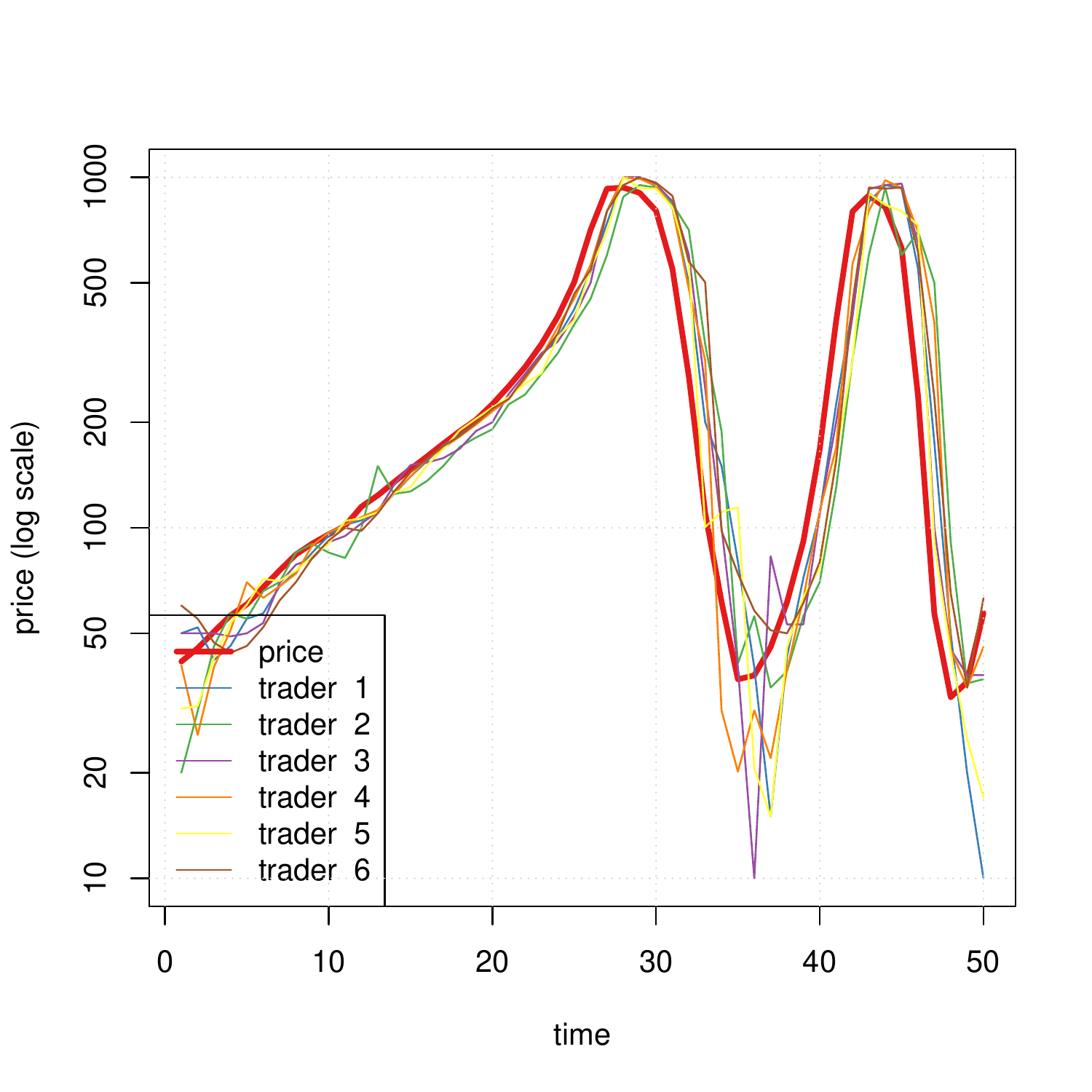}
\caption{%
Price and traders' estimate over time for group 2.
Note that traders' estimates $p^h$ are for time $t+1$ and are used to form the price $p_t$ at time $t$,
i.e.\ $p_t = 1/H (\sum_h p_{t+1}^h + D) / (1+r)$.
}
\label{fig:all1}
\end{figure}

\begin{figure}[ht]
\centering
\includegraphics{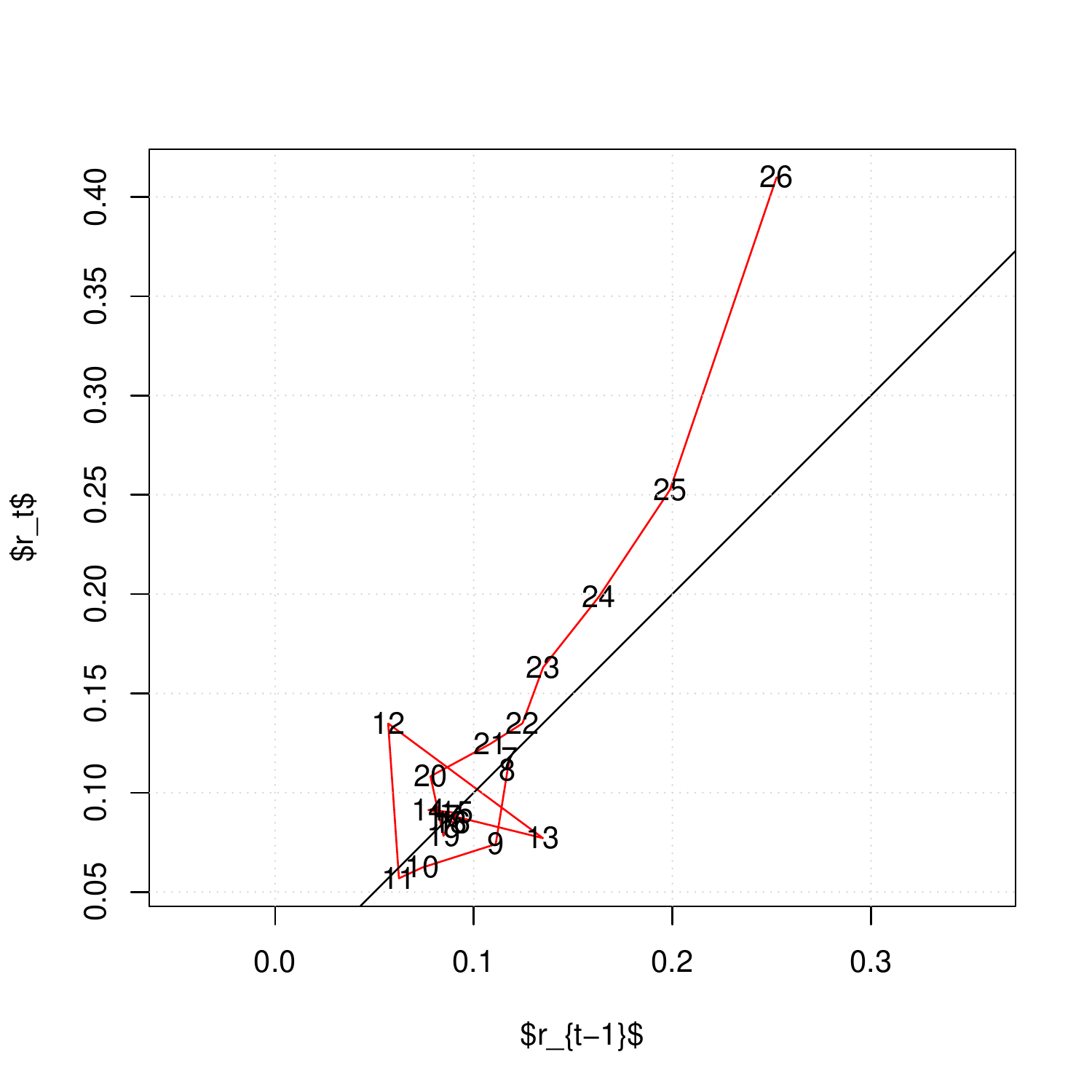}
\caption{%
Next period returns $r_{t+1}$ versus current returns $r_t$ for group 2.
Points on the diagonal correspond to constant growth ($r_{t+1}=r_{t}$),
points above the diagonal ($r_{t+1} > r_{t}$) correspond to accelerating growth.
Note that returns are defined as discrete returns, i.e.\ $r_{t+1} := (p_{t+1} / p_t) -1$.
}
\label{fig:returns1}
\end{figure}

\begin{figure}[ht]
\centering
\includegraphics{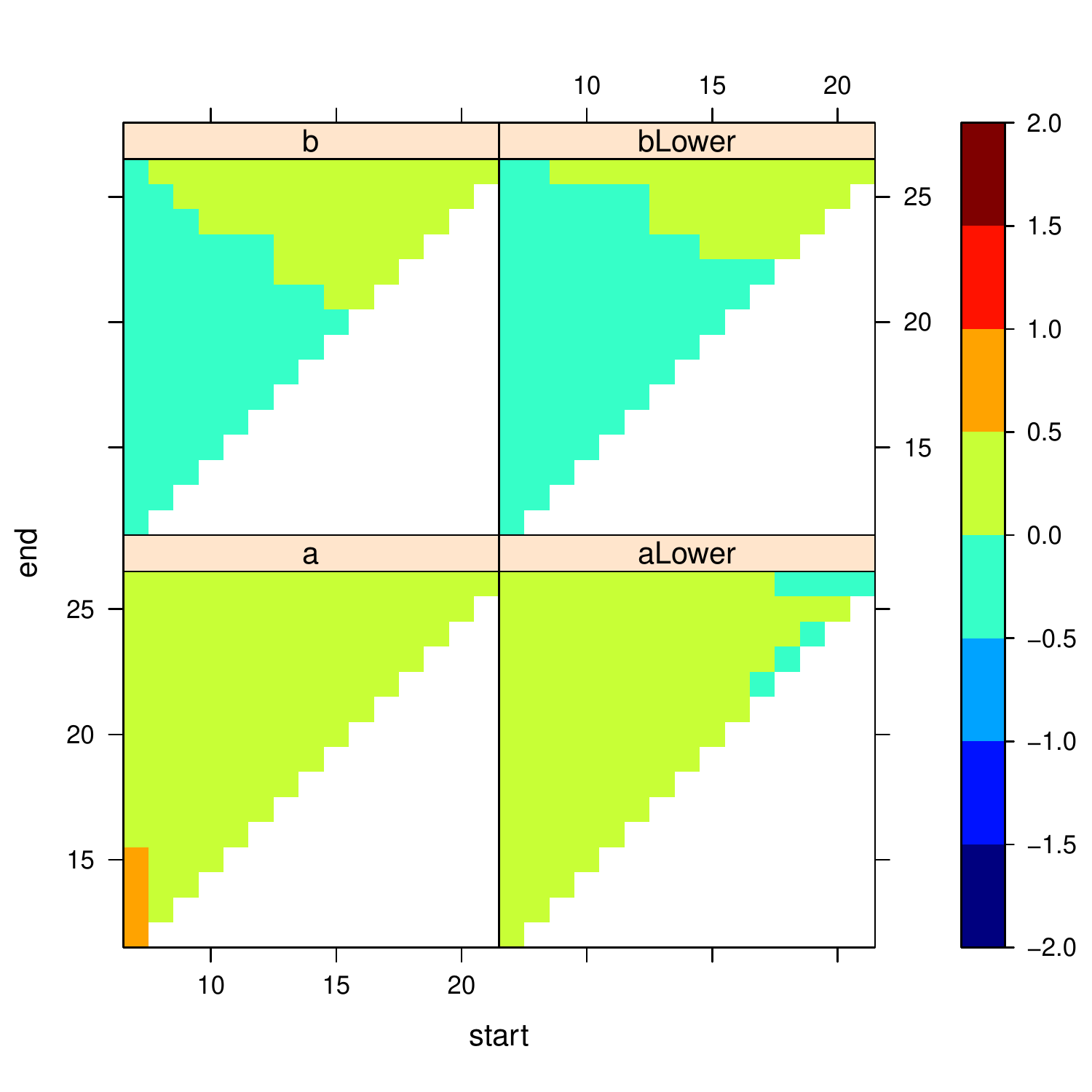}
\caption{%
Parameter estimate of \autoref{eq:regPrice} over the time interval [start, end] for group 2.
The $x$-axis corresponds to the start point and the $y$-axis to the end point of the analyzed time window.\ 
The bar on the right gives the values of the parameters in color code, according to the indicated scale.\
aLower and bLower correspond to the lower 95\% confidence level of $a_2$ and $b_2$ respectively of \autoref{eq:regPrice}.
Note that $b_2$ is around $0$ for small starting and end values implying exponential growth in the initial phase of the bubble.
We observe a rather large domain in the parameter range describing the start time and end time of the window of calibration for
which the parameter $b_2$ is positive at the 95\% confidence level.
}
\label{fig:triangle1}
\end{figure}

The bubble period identified by \cite{Hommes2008} runs from 7 -- 26.
\autoref{fig:all1} shows that the price becomes larger than the fundamental value $p^f$ at $t=7$.
Checking the returns vs.\ past returns in \autoref{fig:returns1},
we see that the bubble initially grows approximately exponentially ($r_t \approx r_{t-1}$) 
as confirmed by the positions of the points along the diagonal.
Later, at around $t=14$, the returns become monotonous increasing (i.e.\  
prices become faster than exponential growth) and 
are plotted above the diagonal.
This is also confirmed by \autoref{fig:triangle1}
where, for low starting and ending values of the analyzing time window,
the parameters estimated for \autoref{eq:regPrice} are not distinguishable from exponential growth since
the parameter $b_2$ is not significantly different from zero.
However, towards the middle and the end of the bubble,
the growth rate accelerates ($b_2$ becomes significantly larger then zero) before the bubble finally bursts.
The parameter $a_2$ is positive over the whole analysis window (lower left panel) and
almost always significantly larger than zero (lower right panel).
The upper panels shows that $b_2$ (for low start and ending values) is not significant different from zero,
but, later in the bubble, $b_2$ becomes positive (top left panel) and even significant positive (upper right panel).

\begin{figure}[ht]
\centering
\includegraphics{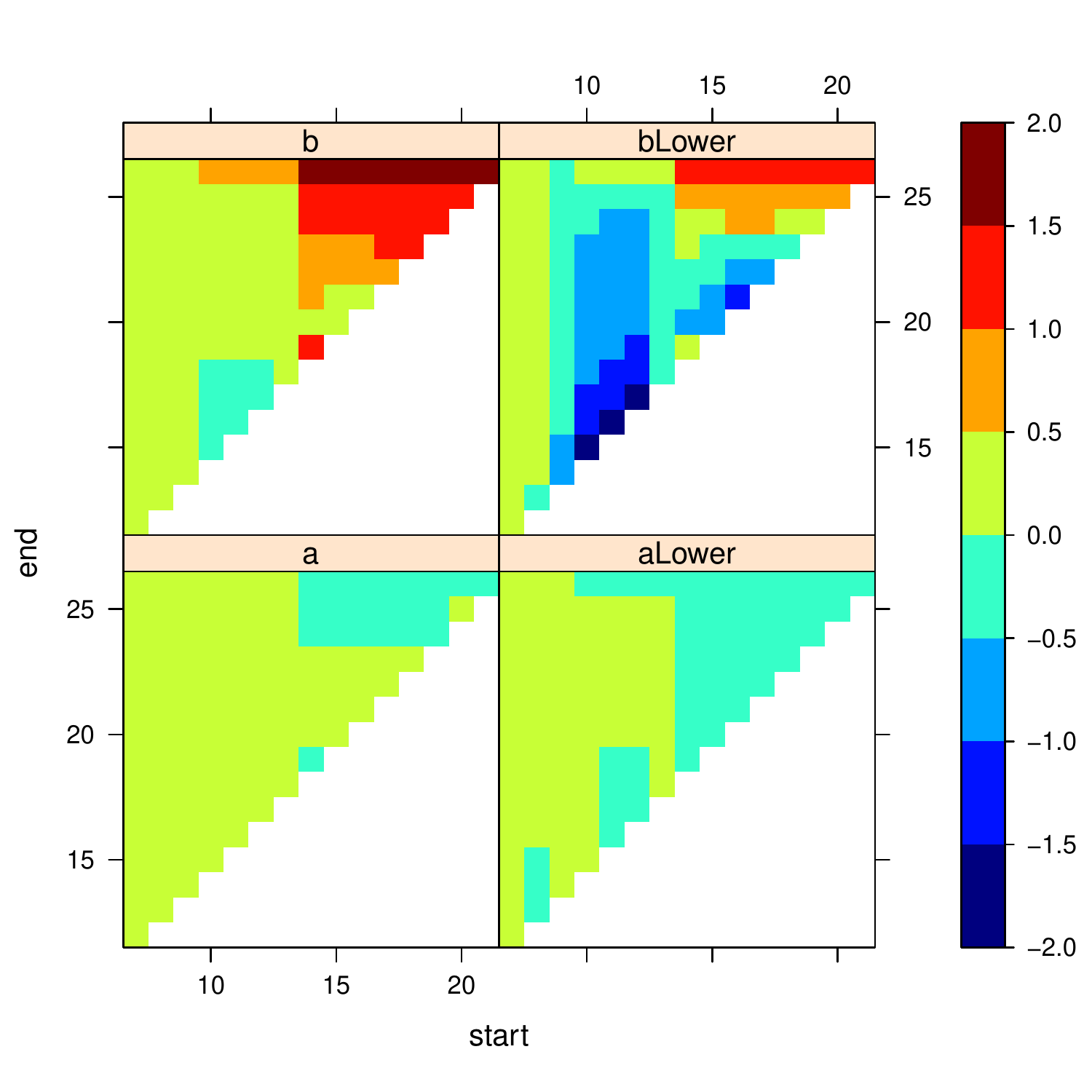}
\caption{%
Parameter estimate of \autoref{eq:regReturn} over the time interval [time, start] for group 2.
The $x$-axis corresponds to the start point and the $y$-axis to the end point of the analyzed time window.\ 
aLower and bLower correspond to the lower 95\% confidence level for $a_3$ and $b_3$ respectively of \autoref{eq:regReturn}.
Note that the domain where $a_3$ and $b_3$ are both significantly larger than zero is restricted to the earliest two starting points.
}
\label{fig:triangle12}
\end{figure}

Checking for the existence of feedback from past returns in \autoref{fig:triangle12}, we find
that \autoref{eq:regReturn} describes less accurately the experimental results;
although the parameters $a_3$ and $b_3$ are both positive (left panels),
the time windows where the parameters are both significantly positive (right panels) is relatively small (only for starting values $t=7$ and $t=8$).

Hence, in summary, the bubble in group 2 does not only grow significantly faster then
exponential in the end phase, but traders seem to anchor their expectations more on price rather than on return.

%%%%%%%%%%%%%%%%%%%%%%%%%%%%%%%%%%%%%%%
\subsection{Group 3}

\begin{figure}[ht]
\centering
\includegraphics{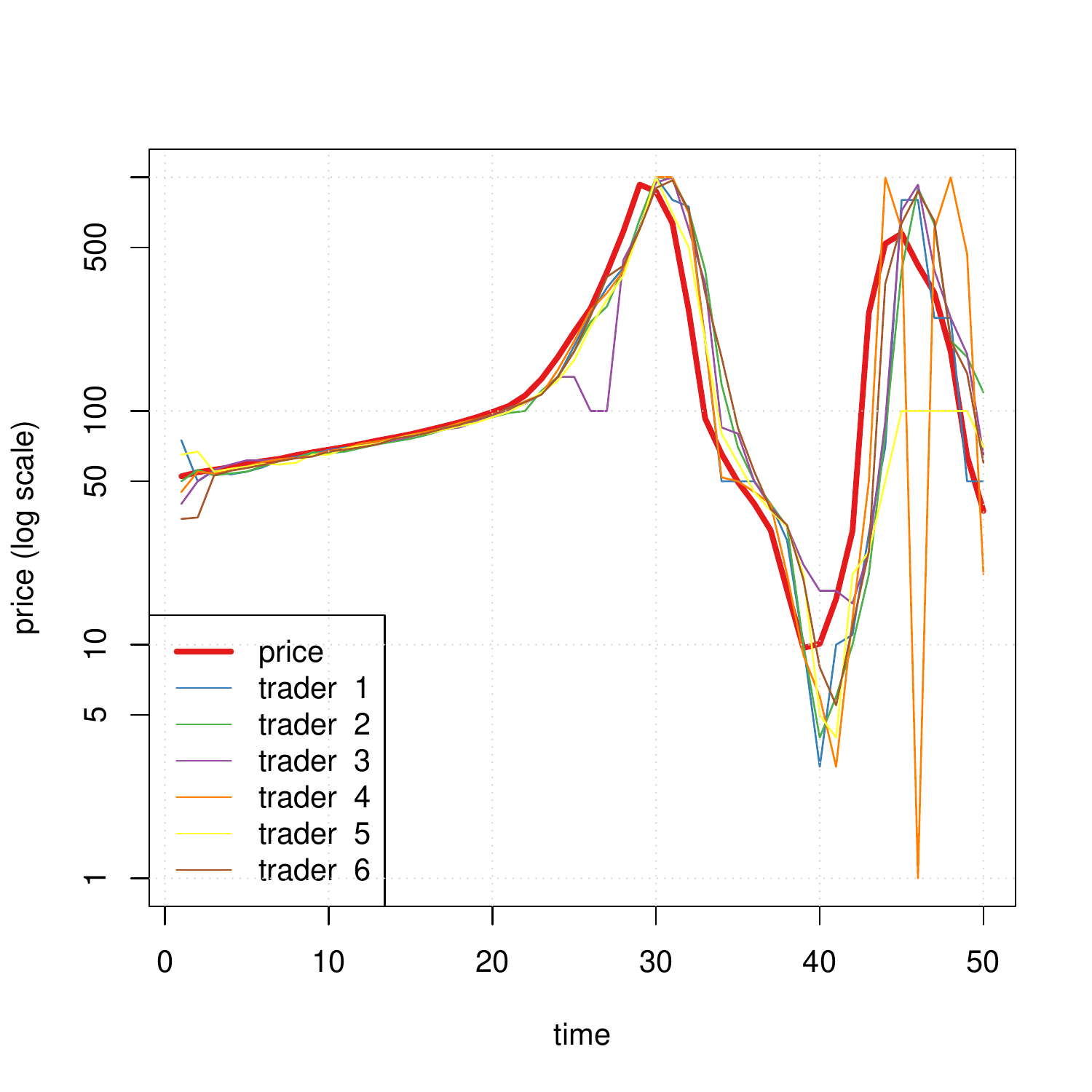}
\caption{Price and traders' estimate over time for group 3. Same representation as \autoref{fig:all1}.}
\label{fig:all2}
\end{figure}

\begin{figure}[ht]
\centering
\includegraphics{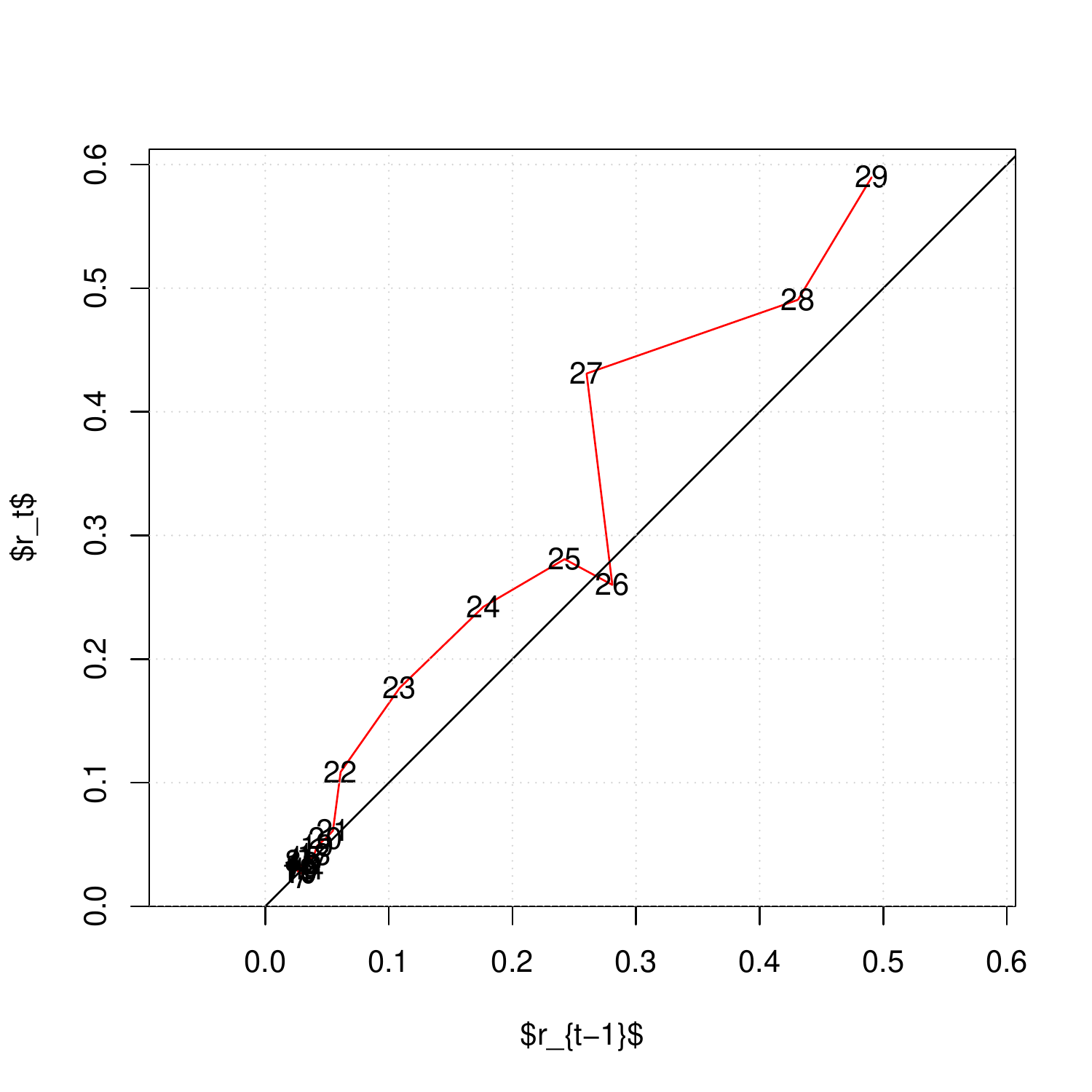}
\caption{Next period returns $r_{t+1}$ versus current returns $r_t$ for group 3. Same representation as \autoref{fig:returns1}.}
\label{fig:returns2}
\end{figure}

\begin{figure}[ht]
\centering
\includegraphics{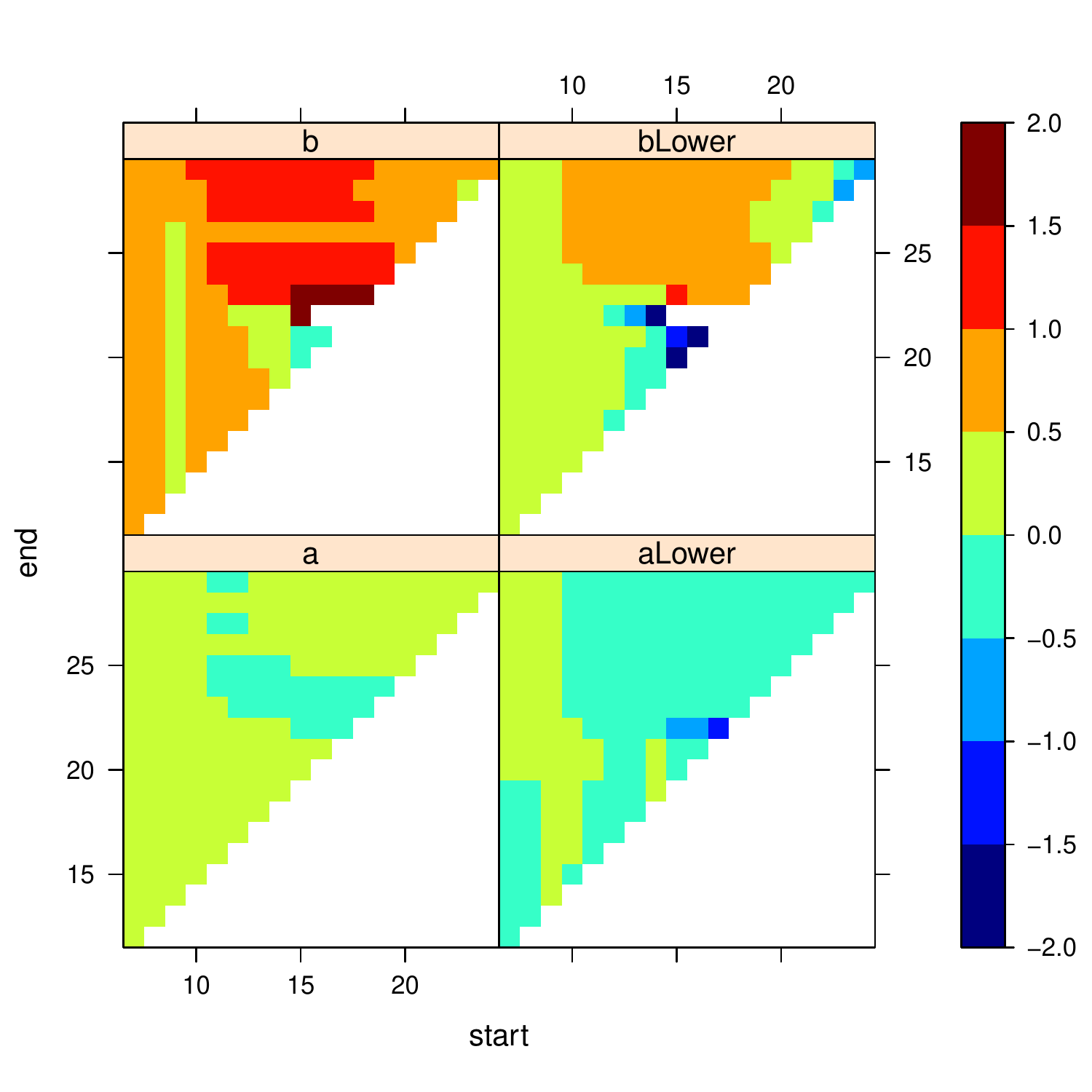}
\caption{Parameter estimate of \autoref{eq:regReturn} over the time interval [time, start] for group 3. Same representation as \autoref{fig:triangle12}.}
\label{fig:triangle22}
\end{figure}

Group 3 (over the time horizon from 7 -- 29) is the longest bubble among the six groups.
From \autoref{fig:all2} (which is plotted on $\log$ scale), the bubble seems to grow initially only exponentially (visible as a straight line in the plot),
which is also confirmed by \autoref{fig:returns2}, which shows that the growth rate is initially constant.
At around $t=20$, growth accelerates.
This observation is also confirmed by the analysis of \autoref{eq:regPrice}, where $a_2$ is
significant for almost all analysis windows.
But, the positive feedback of the price on the growth rate
embodied by $b_2$ becomes only significant in the later phase of the bubble.
Analyzing this group for the possible
existence of anchoring on return (\autoref{eq:regReturn}) in \autoref{fig:triangle22},
we find that the results are less clear cut:
although $a_3$ and $b_3$ are positive, $a_3$ is not significantly different form zero for starting values after $t=10$.
Hence, we conclude that \autoref{eq:regReturn} does not appropriately describe the price and
traders tend to anchor their expectations on price rather than on return.

\begin{figure}[ht]
\centering
\includegraphics{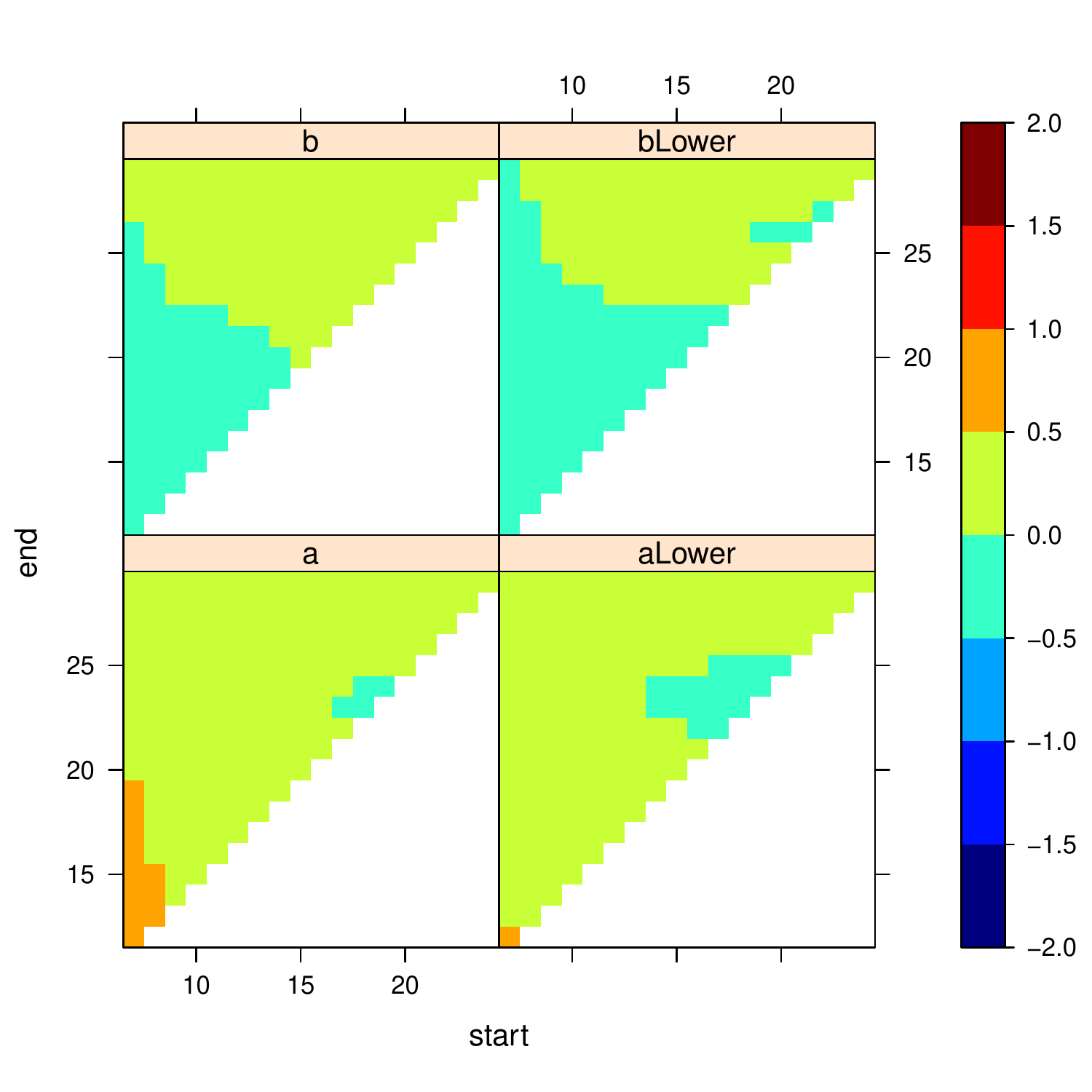}
\caption{Parameter estimate of \autoref{eq:regPrice} over the time interval [start, end] for group 3. Same representation as \autoref{fig:triangle1}.}
\label{fig:triangle2}
\end{figure}

%%%%%%%%%%%%%%%%%%%%%%%%%%%%%%%%%%%%%%%
\subsection{Group 4}
\begin{figure}[ht]
\centering
\includegraphics{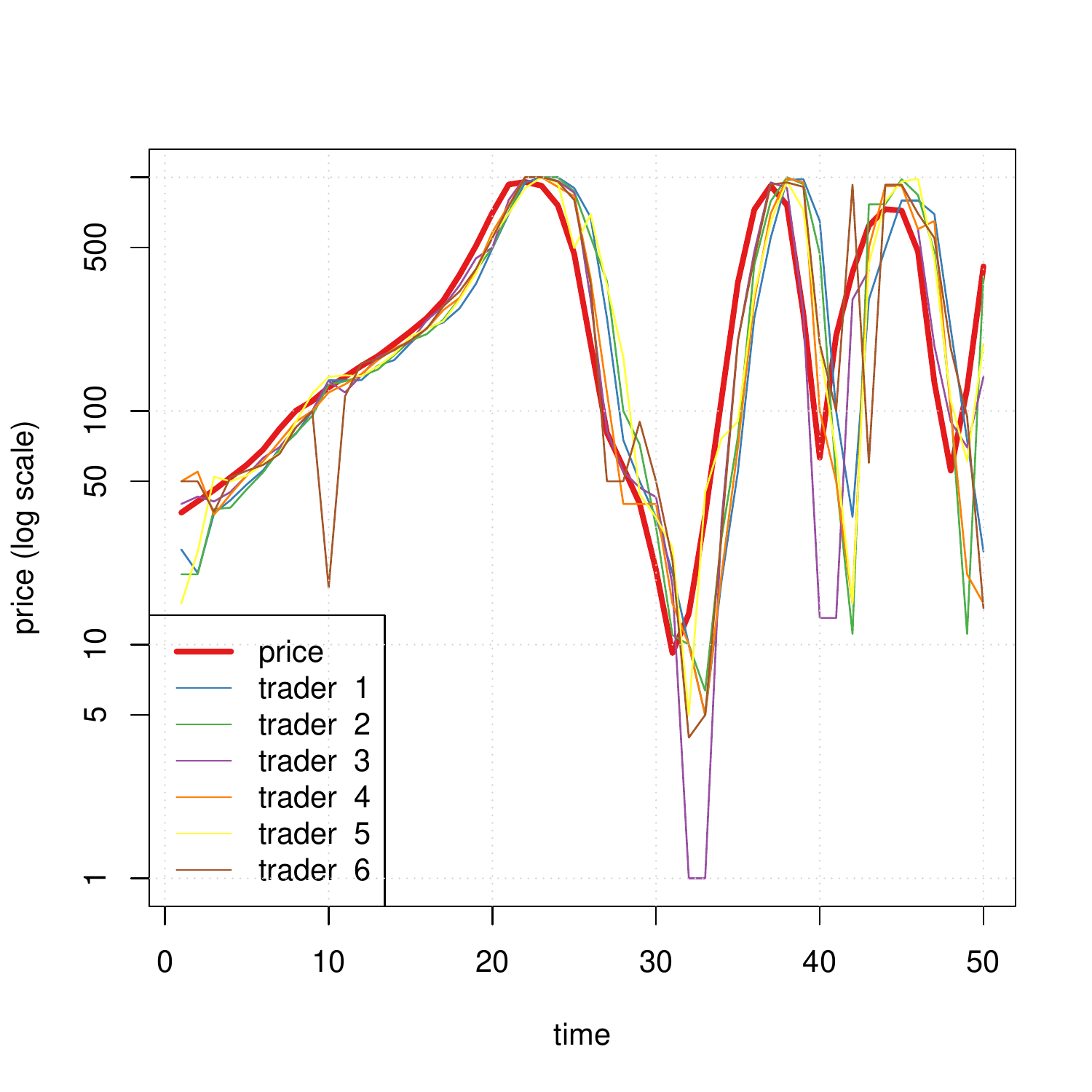}
\caption{Price and traders' estimate over time for group 4. Same representation as \autoref{fig:all1}.}
\label{fig:all3}
\end{figure}

\begin{figure}[ht]
\centering
\includegraphics{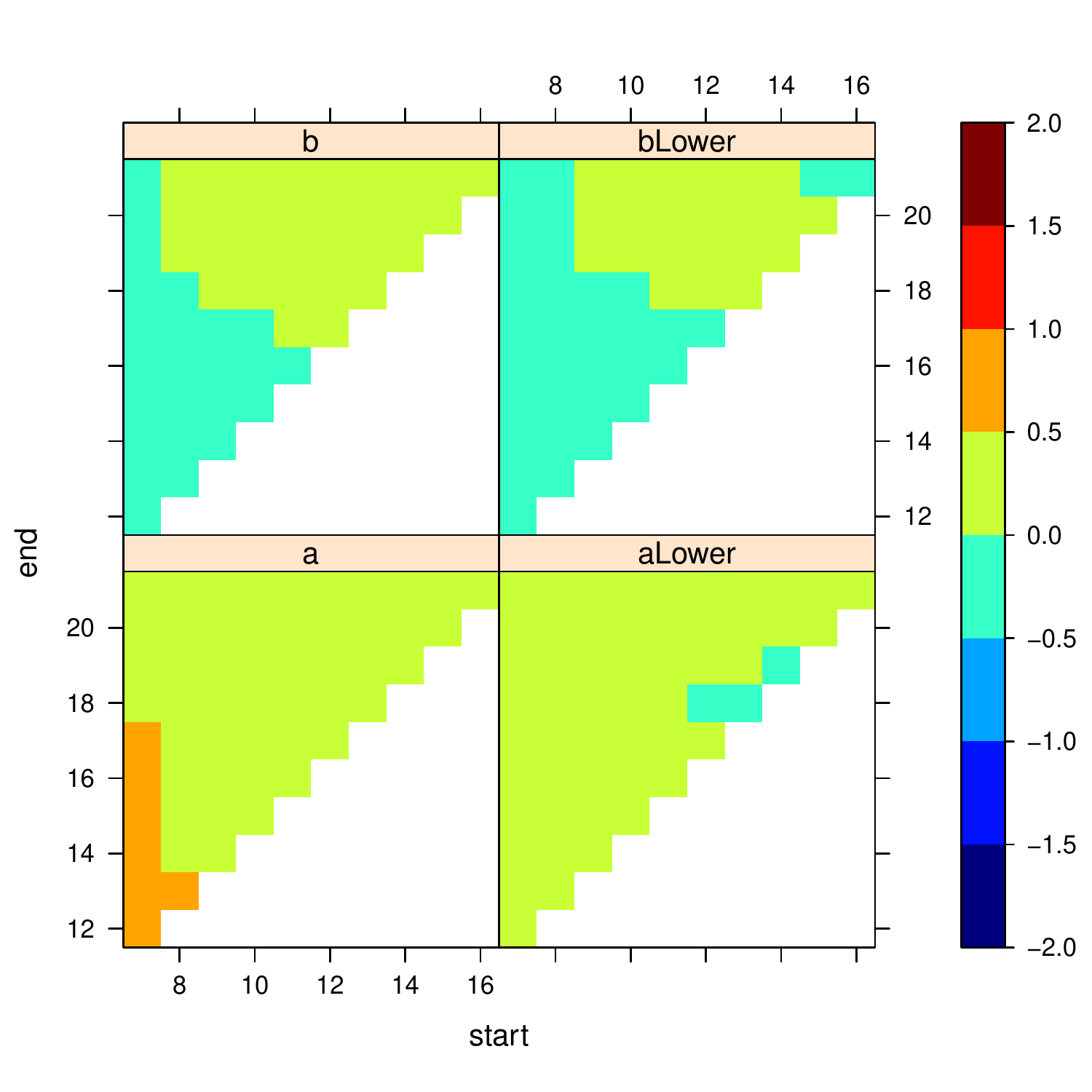}
\caption{Parameter estimate of \autoref{eq:regPrice} over the time interval [start, end] for group 4. Same representation as \autoref{fig:triangle1}.}
\label{fig:triangle3}
\end{figure}

\begin{figure}[ht]
\centering
\includegraphics{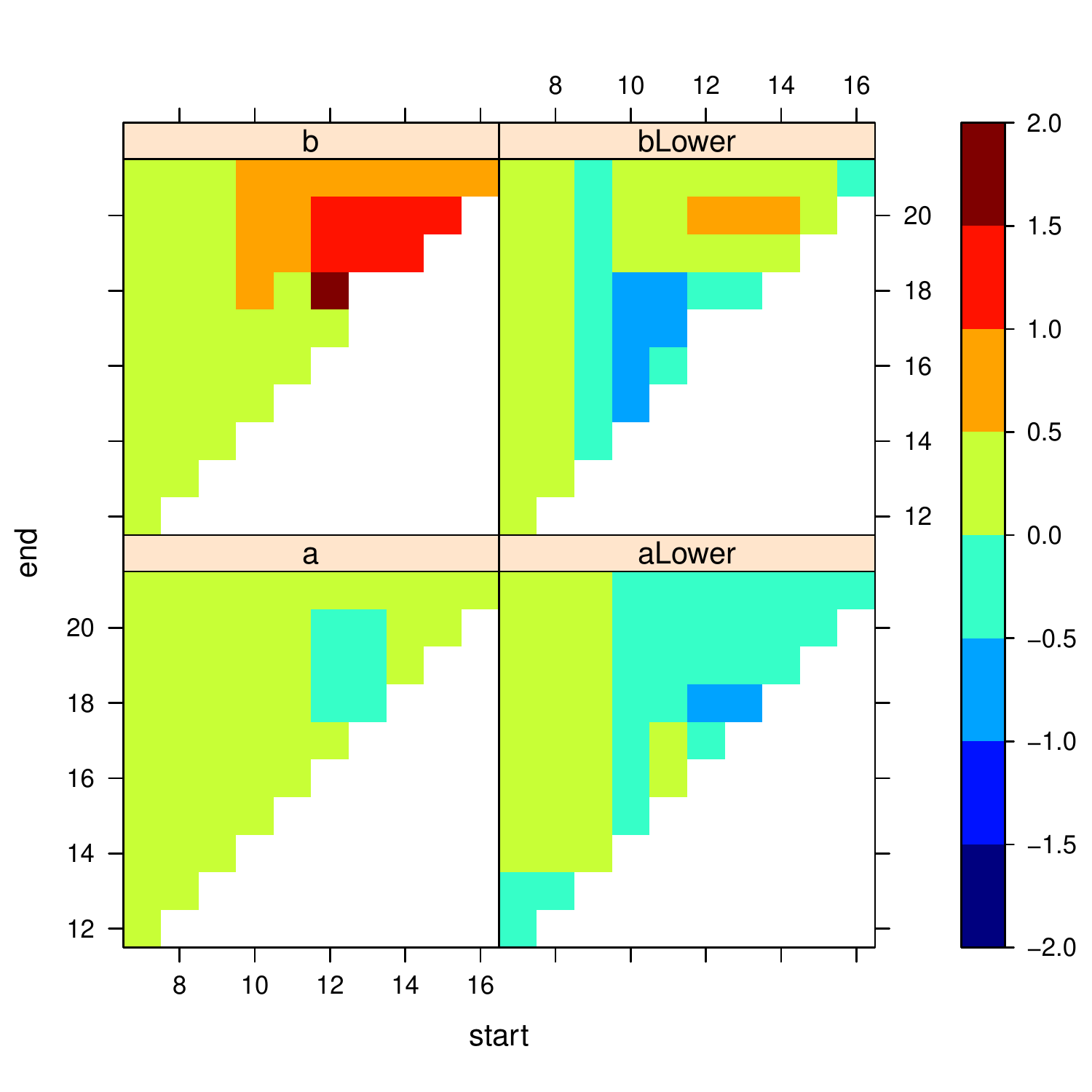}
\caption{Parameter estimate of \autoref{eq:regReturn} over the time interval [time, start] for group 4. Same representation as \autoref{fig:triangle12}.}
\label{fig:triangle32}
\end{figure}

As can be seen from \autoref{fig:all3}, the bubble formed over the time window 7 -- 29 is briefly disrupted by the intervention of trader number 6%
\footnote{The prediction of trader number six at time point $t=10$ seems to be off by an order of magnitude as he has misplaced the decimal.}.
This can also be seen in \autoref{fig:returns2} where we plot the returns.
Between $t=7$ and $t=13$, we have more or less a cobweb and then, starting with $t=14$, the growth rate increases and a bubble is formed.
For anchoring on price, we see in \autoref{fig:triangle3} very strong evidence for faster then exponential growth;
$a_2$ and $b_2$ are both significantly positive.
Again, for very early and small analysis windows, only $a_2$ is positive, indicating exponential growth in the initial phase of the bubble.
The analysis for \autoref{eq:regReturn} in \autoref{fig:triangle32} is less clear,
but the signal for jointly positive $a_3$ and $b_3$ is relatively small (only for two smallest starting values),
indicating that traders prefer to anchor their predictions on price and not on return.

\begin{figure}[ht]
\centering
\includegraphics{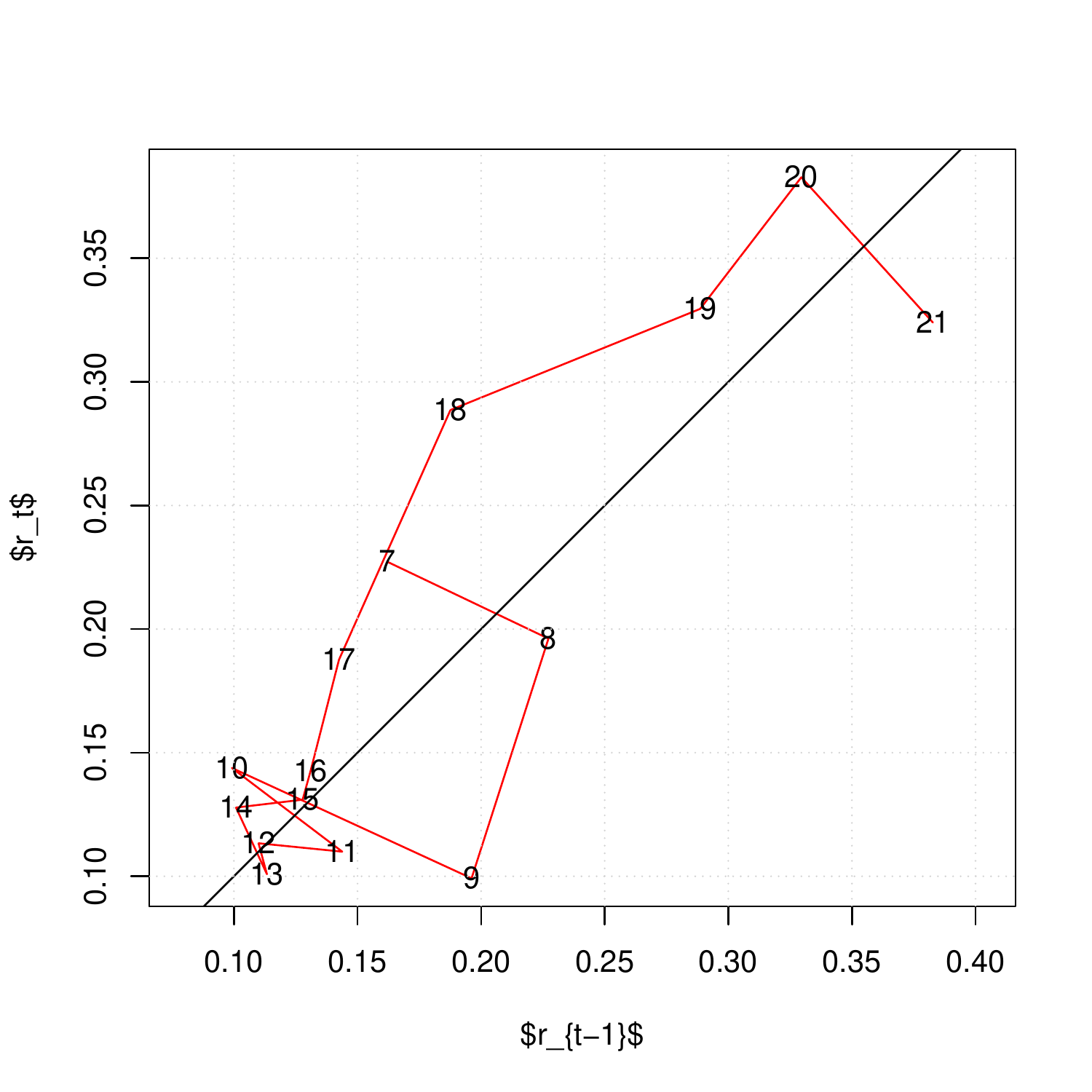}
\caption{Next period returns $r_{t+1}$ versus current returns $r_t$ for group 4. Same representation as \autoref{fig:returns1}.}
\label{fig:returns3}
\end{figure}

%%%%%%%%%%%%%%%%%%%%%%%%%%%%%%%%%%%%%%%%%%%%%%%%%%%%%%%%%%%%%%%%%%%%%%%%%%%%%%%
%This should explore the significance of the results of the work, not repeat them. A combined Results and Discussion section is often appropriate. Avoid extensive citations and discussion of published literature.
\section{Discussion}

It is remarkable that we find many time windows where we can clearly reject the hypothesis of exponential growth and
find evidence for faster than exponential growth.
This is even more remarkable when taking into account that
the data suffers some limitations which make detection of faster than exponential growth more difficult.
\begin{description}
	\item[Price ceiling:]
	Although the price is allowed to fluctuate over a relatively large range, it is capped at
	a maximum value of $1000$.
	Because traders know and can anticipate this,
        we would expect traders to level off  their price expectation much before reaching this upper bound. This turns out
        not to be the case.
	\item[Stable equilibrium price:]
	The pricing formula \autoref{eq:marketPrice} assumes a fundamental value of $60$ and
        thus biases the price towards this value.
	Even if all traders give an estimate of $1000$,
        the realized market price from \autoref{eq:marketPrice} would be $(1000 +3)/1.05 \approx 955$,
        i.e.\ the price is artificially deflated by almost 45 monetary units. 
	\item[Mis-trades:]
	There seems to be a few instances where trades' estimates are off by an order magnitude
        (i.e.\  some traders seem to fail to place the decimal point at the correct digit at some times).
	\item[Short data horizon:]
	Although the experiments run over a time horizon of $50$ time-steps;
        the bubbles appear in much shorter time, leaving relatively few points to estimate tight confidence intervals.
\end{description}

\cite{Heemeijer2009} ran a comparable experiment with a slightly different price forming mechanisms and focusing on the traders' individual price forecasts.
Further, agents' predictions had to lie in a relatively narrow range  (0 -- 100) allowing relatively small deviations from the fundamental price compared to the data that we have analyzed here.
In contrast to \cite{Heemeijer2009} who analyzed the data along the dimensions of trend following, fundamentalism and obstinacy, we focus on non-linear feedback of realized price and return on the price growth rate.
\cite{Anufriev2012} have fitted a heuristics switching model to a positive
feedback asset pricing experiment in the presence of a fundamental robot trader,
whose trading drives the price back towards its fundamental value.
As a consequence, long lasting bubbles do not arise in that setting,
but rather asset prices oscillate around the fundamental and individuals
switch between different simple forecasting heuristics such as adaptive expectations and trend following rules.

\cite{Tirole1982} noted that ``[..] speculation relies on inconsistent plans and is ruled out by rational expectations.''
However, in the experiments of \cite{Hommes2008} that we analyze here, traders are rewarded, not on the basis of how well they predict the fundamental value of the assets they buy but rather, on the accuracy of their prediction of the realized price itself, similarly to real financial markets.
Traders also do not need to invest their wealth into an asset, they do not worry about price fluctuations or care about supply \& demand, which lead them to ``ride the bubble'' (see \cite{Abreu2003}, \cite{DeLong1990} and \cite{DeLong1990B}).
They rather give a forecast as in a Keynesian beauty contest \cite{Keynes1936},
where traders need to synchronize their beliefs.
Such self-confirming predictions can easily lead to herding,
in particular in situations where the fundamental value is not directly observable or
when strong disagreement on the fundamentals between the traders occurs, such as in the dot-com bubbles, see \cite{Shiller2005} for instance.

%%%%%%%%%%%%%%%%%%%%%%%%%%%%%%%%%%%%%%%%%%%%%%%%%%%%%%%%%%%%%%%%%%%%%%%%%%%%%%% 
%The main conclusions of the study may be presented in a short Conclusions section, which may stand alone or form a subsection of a Discussion or Results and Discussion section.
\section{Conclusions}

There have been many reports of super-exponential behavior in financial markets in a literature inspired by
the dynamics of positive feedback leading to finite-time singularities in
natural and physical systems (see for instance \cite{Sornette2001} and \cite{Sornette2004} and
references therein).
However, the challenge has been and is still to confirm with more and more statistical evidence that
the very noisy financial returns do contain a significant positive feedback component during some bubbles regimes.
In the present paper, by analyzing a controlled experiments in the laboratory, we have the luxury of working with a low noise data set.
With this advantage, we have presented the first detailed quantitative calibration of simple models with positive feedback that
unambiguously demonstrate the existence of positive feedback mechanisms in the price formation process of controlled experimental financial markets.

%%%%%%%%%%%%%%%%%%%%%%%%%%%%%%%%%%%%%%%%%%%%%%%%%%%%%%%%%%%%%%%%%%%%%%%%%%%%%%%
%% Acknowledgment
%\section*{Acknowledgment}
%We acknowledge \ldots

\newpage
\appendix

\section*{Appendix}

Faster than exponential growth means
that there is a positive feedback loop, or as \cite{Andreassen1990} noted that
``[..] subjects were more likely [..] to buy as prices rose [..]''.
The table down-below illustrates the difference between constant growth and positive feedback.
Note that the prices in the two bubbles can be indistinguishable in the early phase of the bubble.

\begin{table}[ht]
\begin{center}
\begin{tabular}{c|c c|c c}
{} & \multicolumn{2}{c|}{$\log( \bar{p}_t / \bar{p}_{t-1}) = a_1$}  & \multicolumn{2}{c}{$\log( \bar{p}_t / \bar{p}_{t-1}) = a_2 + b_2 \bar{p}_{t-1}$}  \\ 
$t$ & $\bar{p}_t$& \%   & $\bar{p}_t$& \% \\ \hline
0 & 60.00 &  --  & 60.00 & -- \\ 
1 & 66.00 & 10\% & 65.79 & 10\% \\  
2 & 72.60 & 10\% & 72.19 & 10\% \\  
3 & 79.86 & 10\% & 79.26 & 10\% \\  
4 & 87.85 & 10\% & 87.08 & 10\% \\  
5 & 96.63 & 10\% & 95.74 & 10\% \\  
6 & 106.29 & 10\% & 105.36 & 10\% \\  
7 & 116.92 & 10\% & 116.06 & 10\% \\  
8 & 128.62 & 10\% & 127.99 & 10\% \\  
9 & 141.48 & 10\% & 141.30 & 10\% \\  
10 & 155.62 & 10\% & 156.21 & 11\% \\  
11 & 171.19 & 10\% & 172.95 & 11\% \\  
12 & 188.31 & 10\% & 191.80 & 11\% \\  
13 & 207.14 & 10\% & 213.11 & 11\% \\  
14 & 227.85 & 10\% & 237.30 & 11\% \\  
15 & 250.63 & 10\% & 264.87 & 12\% \\  
16 & 275.70 & 10\% & 296.45 & 12\% \\  
17 & 303.27 & 10\% & 332.86 & 12\% \\  
18 & 333.60 & 10\% & 375.09 & 13\% \\  
19 & 366.95 & 10\% & 424.48 & 13\% \\  
20 & 403.65 & 10\% & 482.74 & 14\% \\  
21 & 444.01 & 10\% & 552.22 & 14\% \\  
22 & 488.42 & 10\% & 636.09 & 15\% \\  
23 & 537.26 & 10\% & 738.87 & 16\% \\  

\end{tabular}
\end{center}
\caption{%
Table illustrating the difference between exponential growth ($a_1=\log(1.1) \approx 0.095$, second column)
and positive feedback by price on future returns  ($a_2=\log(1.09) \approx 0.086$, $b_2 = 0.0001$, fourth column).
We let the bubbles start at $\bar{p}_t=60=120-60=p_t-p^f$.
With the parameter above, the excess price $\bar{p}_t$ grows initially at around 10\% at each time step.
In the early phase, the prices grow approximately exponentially (the exponential growth is actually
slightly faster). At time step $t=10$, the bubble with positive feedback of the price on future returns overtakes the exponential growth
benchmark and the growth rate start to accelerate.}
\label{tab:example}
\end{table}

\begin{figure}[htb]
\begin{center}
%trim option's parameter order: left bottom right top
\includegraphics[trim = 1mm 75mm 1mm 70mm, width=0.8\textwidth]{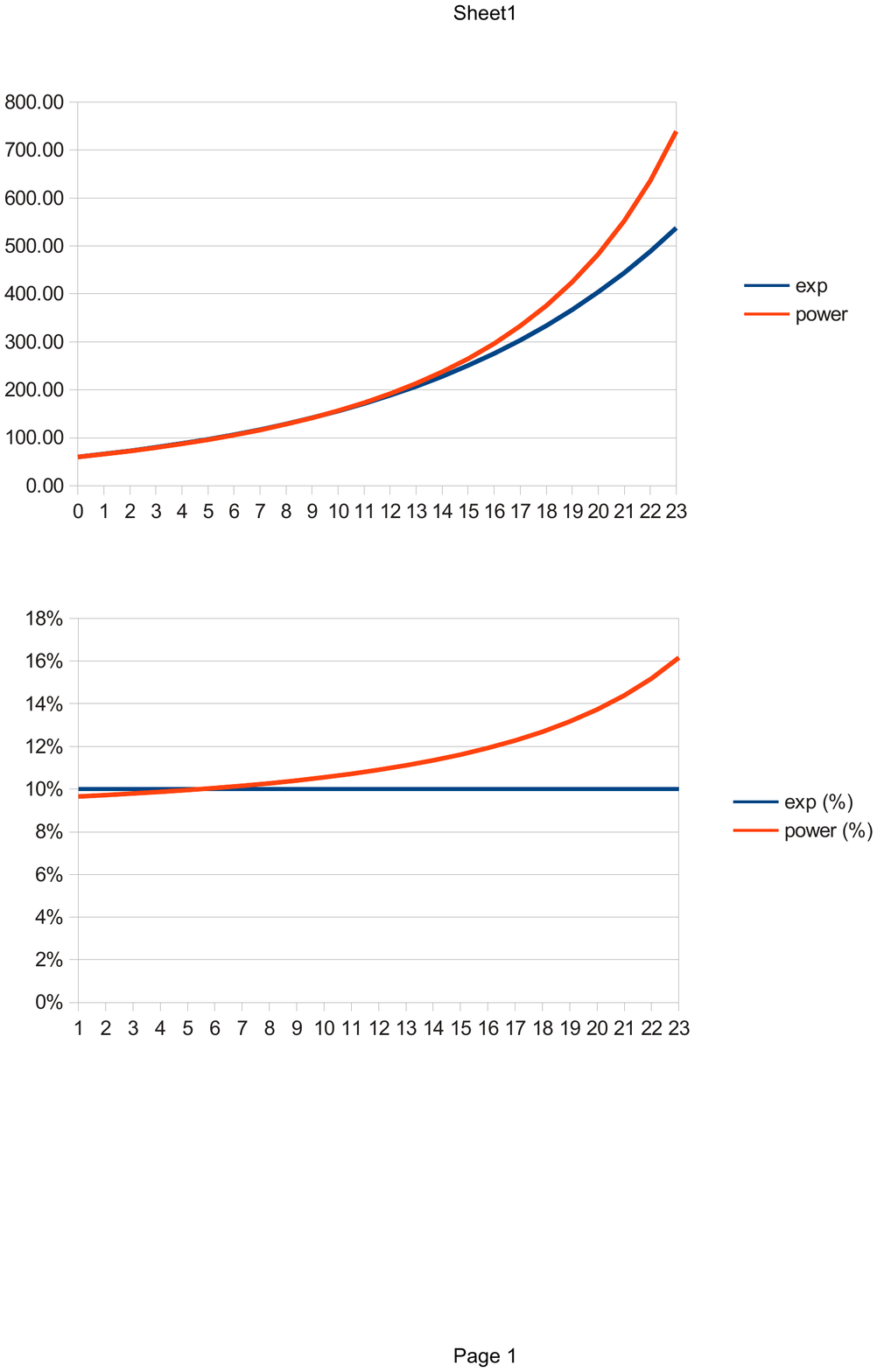}
\caption{Graphical representation of \autoref{tab:example}. Top panel: prices. Bottom panel: returns.}
\end{center}
\label{fig:example}
\end{figure}

\newpage

%%%%%%%%%%%%%%%%%%%%%%%%%%%%%%%%%%%%%%%%%%%%%%%%%%%%%%%%%%%%%%%%%%%%%%%%%%%%%%%
%%%%%%%%%%%%%%%%%%%%%%%%%%%%%%%%%%%%%%%%%%%%%%%%%%%%%%%%%%%%%%%%%%%%%%%%%%%%%%%
%%%%%%%%%%%%%%%%%%%%%%%%%%%%%%%%%%%%%%%%%%%%%%%%%%%%%%%%%%%%%%%%%%%%%%%%%%%%%%%

%% bibtex stuff
%\bibliographystyle{dcu}  % for harvard
\bibliographystyle{abbrvnat}
\bibliography{ref}

\end{document}